\documentclass[showpacs,amssymb,twocolumn,aps]{revtex4}
\usepackage{amsmath}
\usepackage{graphicx} 
\usepackage{epsf}
\usepackage{graphics}
\usepackage[latin1]{inputenc}
\newcommand{\be}{\begin{equation}}
\newcommand{\ee}{\end{equation}}
\newcommand{\bea}{\begin{eqnarray}}
\newcommand{\eea}{\end{eqnarray}}

\begin{document}

\title{Hard Thermal Loops in the $n$-Dimensional $\phi^3$ Theory}

\author{F. T. Brandt  
and J. Frenkel} 
\affiliation{ Instituto de F\'{\i}sica,
Universidade de S\~ao Paulo,
S\~ao Paulo, SP 05315-970, Brazil}

\begin{abstract}
We derive a closed-form result for the leading thermal contributions which appear in the 
\hbox{$n$-dimensional} $\phi^3$ theory at high temperature. These contributions become local only in the long wavelength
and in the static limits, being given by different expressions in these two limits.
\end{abstract}


\maketitle

\section{Introduction}

An important issue in thermal perturbation theory is the study of the high temperature behavior of the Green functions.
These hard thermal loops have simple symmetry properties, but the corresponding effective actions
are in general non-local functionals of the external fields \cite{Braaten:1990mz,Frenkel:1990br}. 
However, in the long wavelength
and in the static limits, the thermal loops yield local functionals 
\cite{Brandt:2009ht,Frenkel:2009pi}. 
The purpose  of this note is to study the behavior of the hard thermal loops in the quantum $\phi^3$ theory
in $n$-dimensions. This theory has several interesting similarities with QCD (for $n=6$) and with quantum
gravity (for $n=8$) \cite{Brandt:2008gy,Brandt:2008cn,Brandt:2012ei}.

It is is well known that the leading thermal contributions which occur in the $\phi^3$ theory at high temperature
appear only in the self-energy functions. Thus, in section II, we derive a general closed-form result for the
two-point function at high temperature $T$ [Eq. \eqref{eq12}]. This result is expressed in terms of a
hypergeometric function, which depends upon the ratio of the external energy and momentum. In section III, we discuss
some consequences of this result. We show that in the long wavelength and static limits, this function reduces to
local expressions, which are independent of the external energy and momentum. 
Nevertheless, these two limits give different results. Moreover, we show that
the leading thermal contributions in the static limit, are the same as  those obtained by setting directly, in the
loops, their external  energy and momentum equal to zero.

\section{The thermal 2-point function}
Let us consider the thermal diagram of Fig.~1. In the imaginary-time formalism 
\cite{kapusta:book89,lebellac:book96,das:book97}, the contributions
of this diagram may be written in the form
\be\label{eq1}
\Pi_{n}(k_0,\vec k) = \frac{1}{2} \frac{\lambda^2}{(2\pi)^{n-1}}\int d^{n-1} Q I,
\ee
where $\lambda$ is the coupling constant and
\be\label{eq2}
I = T \sum_n \frac{1}{Q_0^2-{\vec Q} ^2}\frac{1}{(Q_0+k_0)^2-{\vec   P}^2},
\ee
Here $Q_0 = 2\pi i n T$, $k_0= 2\pi i l T$  ($n,l=0,\pm 1,\pm 2, \dots$) and $\vec P = \vec Q + \vec k$.
The thermal part of $I$ can be expressed as
\be\label{eq2a}
I^T = \frac{1}{2\pi i} \int_C dQ_0 \frac{N(Q_0)}{Q_0^2 - \vec Q^2} \frac{1}{(Q_0+k_0)^2-{\vec   P}^2}, 
\ee
where
\be\label{eq3}
N(Q_0) = \frac{1}{{\rm e}^{Q_0/T} - 1} 
\ee
is the thermal distribution factor and $C$ is the contour which
encloses all the poles of $N$ in an anticlockwise sense. Next, we
evaluate \eqref{eq2a} in terms of the poles outside $C$, to get
\begin{eqnarray}\label{eq4}
 I^T &=& \frac{1}{4 QP} \left\{\left[N(Q) + N(P) + 1\right] 
\right. \nonumber \\ &\times& \left.
\left(\frac{1}{k_0-Q-P}  - \frac{1}{k_0+Q+P} \right)
\right. \nonumber \\ 
&+& \left. 
\left[N(Q) - N(P) \right] 
\right. \nonumber \\ &\times& \left.
\left(\frac{1}{k_0+Q-P}  - \frac{1}{k_0-Q+P} \right)\right\}  ,
\end{eqnarray}
where $Q=|\vec Q|$, $P=|\vec Q + \vec k|$ and we have used 
\be\label{eq5}
N(k_0-Q) = N(-Q) = -N(Q) -1
\ee  
since in the imaginary time formalism $k_0/2\pi i T$ is an integer. Then, after using \eqref{eq5} it is possible to analytically continue $k_0$ to $k_0+i\epsilon$, where now $k_0$ is a continuous real energy.

\begin{figure}[b!]
\label{top1}
  \begin{center}
    \includegraphics[scale=0.3]{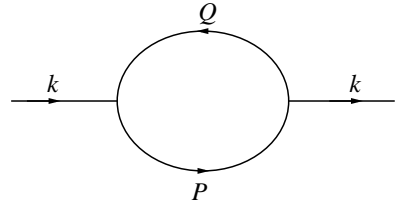}
  \end{center}
\caption{The 2-point self-energy diagram. The internal lines indicate thermal scalar particles.}
\end{figure}

We now look for the leading thermal contribution, which comes from the region 
$Q\simeq T \gg k_0, |\vec k|$, so that we may use the approximation
\be \label{eq7}
P = |\vec Q + \vec k| \simeq Q + \hat{\vec Q}  \cdot \vec k,
\ee
where $\hat{\vec Q} $ is a unit vector in the direction of $\vec Q$. In this way, \eqref{eq4} leads to
\be \label{eq8}
I^T\simeq -\frac{1}{2 Q^2} \left[\frac{N(Q)}{Q} + \frac{N^\prime(Q) \hat{\vec Q} \cdot\vec k}{k_0-\hat{\vec Q} \cdot\vec k}\right].
\ee 
After substituting this result in \eqref{eq1}, we use the integrals \cite{gradshteyn}
\begin{subequations}
\begin{eqnarray} 
\int_0^\infty dQ N(Q) Q^{n-5}  &=& T^{n-4} \Gamma(n-4) \zeta(n-4) \label{eq9a} \\
\int_0^\infty dQ N^{\prime}(Q) Q^{n-4}  &=& -T^{n-4} \Gamma(n-3) \zeta(n-4) \label{eq9b} ,
\end{eqnarray}    
\end{subequations}
where $\Gamma$ is the gamma function and $\zeta$  is the Riemann zeta function. Proceeding in this way, we obtain 
\begin{eqnarray}\label{eq10}
\Pi_n(k_0,\vec k) &=& \frac{T^{n-4} \Gamma(n-4) \zeta(n-4) }{4 (2\pi)^{n-1}} \lambda^2
\nonumber \\ &\times &
\int d\Omega\left[\frac{(n-4) k_0}{k\cdot\hat Q} - (n-3) \right],
\end{eqnarray}
where $\hat Q = (1, \hat{\vec Q})$ and $\int d\Omega$ denotes angular integrations over the $n-1$ dimensional unit vector $\hat{\vec Q}$. We see that, in general, $\Pi_n$ would be a non-local function, which in configuration space would involve angular integrals of the form $(\hat Q\cdot \partial)^{-1}$.

In order to perform the angular integrations in \eqref{eq10}, we use
\be\label{eq11}
\int d\Omega = \frac{2 \pi^{\frac{n-2}{2}}}{\Gamma\left(\frac{n-2}{2}\right)} 
\int_0^\pi d\theta (\sin\theta)^{n-3}  \dots ,
\ee
where $\theta$ is the polar angle between the $n-1$ dimensional vectors ${\vec k}$ and $\hat{\vec Q}$. 
The angular integral in the first term of Eq. \eqref{eq10} can be expressed in terms of hypergeometric functions 
\cite{gradshteyn}. To this end we employ \eqref{eq11} in order to write the integral as
\begin{eqnarray}\label{eq11a}
J_{n-1} &=& \int d\Omega \frac{k_0}{k\cdot Q}  = 
\frac{2 \pi^{\frac{n-2}{2}}}{\Gamma\left(\frac{n-2}{2}\right)} \int_0^\pi d\theta
\frac{(\sin\theta)^{n-3}}{1-\frac{|\vec k|}{k_0}\cos\theta} 
\nonumber \\ &=&
\frac{2 \pi^{\frac{n-2}{2}}}{\Gamma\left(\frac{n-2}{2}\right)} \frac{B\left(\frac 1 2,\frac{n-2}{2}\right)}
{1-\frac{|\vec k|}{k_0}} 
\nonumber \\ &\times&
 F\left(1,\frac{n-2}{2};n-2;\frac{2}{1-\frac{k_0}{|\vec k|}}  \right),
\end{eqnarray}
where $B$ denotes the beta function. With the help of transformations formulas \cite{gradshteyn} for the hypergeometric function $F\left(1,\frac{n-2}{2};n-2;\frac{2}{1-{k_0}/{|\vec k|}}  \right)$, the expression \eqref{eq11a} can be further simplified to
\be \label{eq11b}
J_{n-1} = \frac{2 \pi^{\frac{n-1}{2}}}{\Gamma\left(\frac{n-1}{2}\right)} 
F\left(\frac 1 2,1;\frac{n-1}{2};\frac{|\vec k|^2}{k_0^2}  \right).
\ee
Substituting \eqref{eq11b} into \eqref{eq10}, and using again \eqref{eq11} in the second term of \eqref{eq10}, we obtain for the 2-point function the following closed form result
\begin{widetext}
\begin{eqnarray}\label{eq12}
\Pi_n(k_0,\vec k) = \lambda^2 T^{n-4} \frac{\zeta(n-4)}{2^n \pi^{(n-1)/2} } 
\frac{\Gamma(n-4)}{\Gamma\left(\frac{n-1}{2}\right)}
\left[(n-4)F\left(\frac 1 2,1;\frac{n-1}{2};\frac{|\vec k|^2}{k_0^2}  \right) - (n-3) \right],
\end{eqnarray}
\end{widetext}
which shows that the leading high-temperature contributions in the $n$-dimensional $\phi^3$ theory are proportional to $T^{n-4}$. The closed-form expression \eqref{eq12} reduces to simple functions when $n$ is an integer. For example, for $n=6$, it leads to the well known result
\begin{eqnarray}\label{eq13}
\Pi_6(k_0,\vec k) &=& \frac{\lambda^2 T^2}{96\pi}\left(
1-\frac{k_0^2}{|\vec k|^2}   \right)
\nonumber \\ &\times& 
\left[ 
\frac{k_0}{2|\vec k|}\ln\left(\frac{\frac{k_0}{|\vec k|} + 1}{\frac{k_0}{|\vec k|} - 1}\right)-1\right],
\end{eqnarray}
 which is somewhat analogous to the one obtained for the $\Pi_{00}$ component of the polarization tensor in QCD. Similarly, for $n=8$, we get a $T^4$ behavior at high temperature, which is like that found in thermal quantum gravity \cite{Brandt:1993bj}.

\section{Discussion}
The general result \eqref{eq12} for $\Pi_n(k_0,\vec k)$ is a function of the ratio $|\vec k|/k_0$. In the long wavelength limit, $\vec k =0$, the hypergeometric function has the value $1$, so that we find
\be\label{eq14}
\Pi_n(k_0,\vec k = 0) = - \lambda^2 T^{n-4} \frac{\zeta(n-4)}{2^n \pi^{(n-1)/2} } 
\frac{\Gamma(n-4)}{\Gamma\left(\frac{n-1}{2}\right)} .
\ee
In order to find the behavior of the self-energy $\Pi_n(k_0,\vec k)$ in the static limit $k_0\rightarrow 0$, we may use the transformation formula \cite{gradshteyn}
\begin{eqnarray}\label{eq15}
F\left(\frac 1 2,1;\frac{n-1}{2};\frac{{\vec k}^2}{k_0^2} \right)&=& 
\frac{\Gamma\left(\frac{n-1}{2}\right) \Gamma\left(\frac 1 2\right)}{\Gamma\left(\frac n 2 -1\right)}
\left(-\frac{k_0^2}{|\vec k|^2}\right)^{\frac 1 2}
\nonumber \\ &\times& 
F\left(\frac 1 2,2-\frac n 2;\frac{1}{2};\frac {k_0^2}{{\vec k}^2} \right)
\nonumber \\ &+& 
\frac{\Gamma\left(\frac{n-1}{2}\right) \Gamma\left(-\frac 1 2\right)}{\Gamma\left(\frac{n-3}{2}\right)
\Gamma\left(\frac 1 2\right)}
\left(-\frac{k_0^2}{|\vec k|^2}\right)
\nonumber \\ &\times& 
F\left(1 ,\frac{5-n}{2};\frac{3}{2};\frac {k_0^2}{{\vec k}^2} \right),
\end{eqnarray}
which shows that the hypergeometric function in \eqref{eq12} actually vanishes at $k_0=0$. Thus, in the static limit, we get that
\begin{eqnarray}\label{eq16}
\Pi_n(k_0=0,\vec k) &=& - (n-3)\lambda^2 T^{n-4} 
 \nonumber \\ &\times& \
\frac{\zeta(n-4)}{2^n \pi^{(n-1)/2} } 
\frac{\Gamma(n-4)}{\Gamma\left(\frac{n-1}{2}\right)} .
\end{eqnarray}
We can see from \eqref{eq14} and \eqref{eq16}  that 
these two expressions are, in fact, independent of the external
energies and momenta. However, unless $n=4$, 
the results obtained in the long wave and in the static limit are, in general, different.

Finally, let us compare the static result with the one obtained by setting, from the start, 
$k_0=0$ and $\vec k=0$ in the Feynman diagram in Fig 1. This gives (see Eq. \eqref{eq2a})
\be\label{eq17}
I^T(k_\mu=0) = \frac{1}{2\pi i} \int_C dQ_0 N(Q_0)\frac{1}{(Q_0^2-(\vec Q)^2)^2}.
\ee
We can evaluate the $Q_0$ integral by residues, in terms of the double poles outside $C$. Using \eqref{eq5}, we can express the contributions from the double pole at $Q_0=-Q$ in terms of the one at $Q_0=Q$. Then, omitting a $T$-independent term, we readily obtain the result
\begin{eqnarray}\label{eq18}
I^T(k_\mu=0) &=& 2 \frac{d}{dQ_0}\left[\frac{N(Q_0)}{(Q_0+Q)^2}\right]_{Q_0=Q} 
\nonumber \\ &=&
-\frac{1}{2Q^2}\left[\frac{N(Q)}{Q} - N^\prime(Q)\right],
\end{eqnarray}
which coincides with that in \eqref{eq8}, evaluated in the static limit $k_0=0$. Actually, such 
an agreement is expected on general grounds \cite{Frenkel:2009pi}, because the static limit is the only case when analytic continuation does not modify the original thermal distribution function. This may also be seen from Eq. \eqref{eq5}, where the relation 
\hbox{$N(k_0-Q)=N(-Q)$} would no longer hold after analytic continuation, unless $k_0=0$.

In conclusion, we point out that the analysis presented in this work may be 
readily extended to the case of $n$-dimensional 
QED at high temperature. This investigation will be reported elsewhere.

\acknowledgments

We would like to thank CNPq (Brazil) for a grant.


\end{document}